# RETHINKING INDONESIA'S PUBLIC DEBT IN THE ERA OF NEGATIVE INTEREST RATE-GROWTH DIFFERENTIALS

Mervin Goklas Hamonangan[1]

1. Departemen Ilmu Ekonomi, Fakultas Ekonomi dan Bisnis, Universitas Indonesia, Depok, 16424

Email: *mervin.goklas@ui.ac.id*

## ABSTRACT

This study contributes to the discussion about how higher public debt may not be costly because of the negative interest rate-growth differentials by simulating OLG models introduced by Blanchard (2019) under uncertainty, showing debt and welfare dynamics in two scenarios: intergenerational transfers and debt rollovers in the case of Indonesia. The simulation is done by modifying the model parameters based on interest rate-growth differentials historic data from 2004-2019. It is found that the fiscal consensus does not hold when implementing Blanchard's (2019) analysis with Indonesian-based rate parameters. Increasing public debt makes the economy more volatile and high risk. Modifying other factors supports the initial finding, with lower initial endowment diminishing the benefits of public debt and higher capital share under Cobb-Douglas. When the threat of debt explosion appears, efforts to reduce debt share will reduce the welfare of the society. The policy implication is to be careful of the opportunity; increasing public debt may not be the way to go, avoiding possible dire consequences.

# MEMPERTIMBANGKAN KEMBALI UTANG PUBLIK INDONESIA DI ERA SELISIH NEGATIF ANTARA TINGKAT SUKU BUNGA DAN PERTUMBUHAN EKONOMI

## ABSTRAK

Studi ini berkontribusi pada diskusi tentang bagaimana utang publik yang lebih tinggi mungkin tidak mahal karena selisih negatif antara suku bunga dan pertumbuhan ekonomi dengan mensimulasikan model OLG yang diperkenalkan oleh Blanchard (2019) di bawah ketidakpastian, menunjukkan dinamika utang dan kesejahteraan dalam dua skenario: transfer antargenerasi dan perputaran utang dalam kasus Indonesia. Hal ini dilakukan dengan modifikasi parameter model berdasarkan data historis selisih suku bunga dan pertumbuhan ekonomi tahun 2004-2019. Ditemukan bahwa konsensus fiskal tidak berlaku ketika menerapkan analisis Blanchard (2019) dengan parameter berbasis Indonesia. Meningkatnya utang publik membuat perekonomian semakin bergejolak dan berisiko tinggi. Memodifikasi faktor-faktor lain mendukung temuan awal, dengan dana abadi yang lebih rendah mengurangi manfaat utang publik, serta persentase modal yang lebih tinggi di bawah Cobb-Douglas. Ketika risiko ledakan utang muncul, upaya untuk mengurangi utang akan mengurangi kesejahteraan masyarakat. Implikasi kebijakannya adalah berhati-hati terhadap peluang; meningkatkan utang publik mungkin bukanlah cara yang tepat, menghindari kemungkinan konsekuensi yang mengerikan.

**Kata kunci**: OLG, utang publik, selisih suku bunga-pertumbuhan, kesejahteraan, bergejolak.

# 1. INTRODUCTION

The natural interest rate is in decline and among advanced economies projected to stay low (Liu et al., 2022). Significant challenges have arisen for the policymakers, including the issue of secular stagnation, the threat of asset price bubbles and monetary policies where the zero lower bound restricts interest rate policy (A. R. Mian et al., 2021). When the economy is in recession, the central bank may not be able to sufficiently lower the policy rate to support economic recovery (Hofmann et al., 2021). A possible explanation of this decline is the recent theory of indebted demand which finds that the natural interest rate is decreasing because rising income inequality and market liberalization have resulted in indebted household demand. Contrary to the usual view of the decline in r*, data on saving behavior in the United States shows how growing economic disparity is more relevant than demographic trends caused by baby boom generation aging (A. Mian et al., 2021).

Simultaneously, interest rate-growth differential (*r-g*) has declined since the 1980s – and after the 2008 Global Financial Crisis, it turned negative for several countries (Lian et al., 2020). Looking at both trends, Blanchard (2019) proposed several arguments based on his findings that debt might not have a significant financial consequence and only cause limited welfare costs, countering the widespread advocacy to decrease government debt levels for their high levels and consequently unsustainable. Pursuing fiscal expansion would be the way to boost economic growth since negative long-run *r-g* episodes signify better debt sustainability, and low-interest rates environment makes countercyclical fiscal measure perhaps more helpful than its alternatives (Furman & Summers, 2020). Nevertheless, the future is still uncertain. Rogoff (Blanchard et al., 2021) points out that the situation may not be the norm, and instead, we are currently facing a financial crisis with very long overhang effects; at some point, r might go up. Recent literature indicates that Blanchard's view may have been overly optimistic; there are still limits to how much government debt can go (Barrett, 2018; Mauro & Zhou, 2021; Cline, 2021).

In his upcoming book, Blanchard (2023) is fully aware of the criticisms of his general idea of public debt and low interest rates. The approach may seem intimidating, with an oversimplification of infinite fiscal space for the government by only looking at simple arithmetics. It should be noted that higher debts increase aggregate demand and eventually neutral rate r*, reducing the growth differential, and uncertainty occurs regarding the probability of debt explosion (unsustainable debt). Investor behavior is one of the factors contributing to the mentioned probability, creating what is called "sunspot equilibria". To remove the possibility of multiple equilibria, extremely low debt levels are required, substantially below existing debt levels. Realistic debt reductions over the next few decades will not be enough to eliminate this risk.

The declining trend of interest rates is not restricted to only Advanced Economies (AE). Several Emerging Asian economies have seen their natural interest rates plummet during the last two to three decades (Tanaka et al., 2021). At a glance, this evidence may imply that the fiscal consensus[1] can be implemented in the context of AEs and in the Emerging Market (EM). According to Blanchard et al. (2021), this is not the case. They explore the fiscal consensus under the context of India as one of the EM and consider relevant data on four other EMs, including Indonesia. Debt limits are tighter in EMs, where growth uncertainty and interest rates are higher -- fiscal adjustment options are more constrained.

Emerging Markets include ASEAN as a significant part. ASEAN countries have a combined GDP of more than the world's fifth largest economy (ASEAN Secretariat, 2020). Indonesia generates 35.4% of ASEAN's GDP, making it the largest economy in the region (see figure below) and the world's 16th largest, while being the only G20 member state in Southeast Asia and the world's 4th largest population. The state of Indonesia's economy matters to the world, making it an interesting case to dive into.

Overall, the current state of literature is still in the exploratory and development phase, filled with high-engagement debates among many economists. Many uncertainties are involved, and potential gaps to be understood and explored. Quite surprisingly, the literature still lacks the involvement of Emerging Markets in the analysis, focusing only on the AEs. This study aims to fill that gap, using the data on Indonesia's interest-growth differentials. Specifically, how is the possibility of Indonesia utilizing the opportunity of negative *r-g* differentials?

In the result part, exploratory analysis using stylized facts will be done for Indonesia to understand EM debt capabilities with this newfound, still highly debated consensus. Current policies and economic indicators serve as the basis to conclude the sustainability of debt and replicate Blanchard's model on public debt sustainability and low rates. Indonesia's position and ideal policy considering the new fiscal consensus will be assessed. Without changing the two-period OLG model, different parameters fulfill the purpose. This explains the possibility of Indonesia, as an emerging economy with uncertainty and low risk of facing the future economic Zero Lower Bound (ZLB), making use of the *r-g* differential.

---

[1] The new fiscal consensus are made of three considerations: first, because private sector demand has been chronically weak, macro policy actions are required to improve aggregate demand to meet supply; second, because monetary policy instruments have largely been exhausted, fiscal policy should be the primary macro tool for closing the production gap; third, there is room to utilize fiscal policy in this way because, despite its high level, government debt appears to be manageable. These propositions have been increasingly embraced by economists and policymakers, both in the United States and in Europe.

The broad implementation of the consensus affects the economy, and the possibility of the underlying concepts analyzed in the context of EMs, specifically Indonesia, shall make the discussion pose significance and contribute to the growing literature on the topic. Based on the results, conclusions are made as considerations to rethink Indonesia's public debt policy. Will the implementation of fiscal consensus impact how fiscal policy, specifically on the scope of public debt, should be conducted?

**2. LITERATURE REVIEW**

*2.1. Trend of Low Interest Rates*

Wicksell (1936), who linked the marginal product of capital with the interest rate (cost of borrowing the money), postulated that when the cost of borrowing is less than the return on capital, businesses will borrow to buy capital (buildings and equipment), hence raising prices and demand. He defines the natural rate as the one that is consistent with stable inflation and output remaining at its potential level. Several works have concluded that the declining trend will be here to stay (Gourinchas & Rey, 2019; Rachel & Summers, 2019; Kiley, 2020).

Broad literature explains its causes, with little to no consensus regarding which factors contribute the most. This includes a global savings glut (Bernanke, 2005), declining investment rates for the lack of investment opportunities (Gordon, 2012), rising demand for safe investments (Caballero et al., 2021), and a severe financial crisis has thrown a long shadow; thus the deleveraging process is taking its time. (Lo & Rogoff, 2015). Marx et al. (2021) use a calibrated OLG model to analyze why interest rates have been falling, which is not the case for return on capital. The answer is the increase in risk premia, partly because bonds have become a better hedge for stocks and higher risk aversion among investors. A. R. Mian et al. (2021) explain that the rising income inequality is more important than the demographic change occurrences caused by the baby boom generation aging.

*2.2. Debt and Interest Rates*

Government debt is inextricably related to the viability of fiscal deficit. (Brunnermeier et al., 2022). Debt is sustainable in deterministic models, and if the risk-free interest rate *r* is smaller than the economic growth rate *g*, debt rollover is possible. Bohn (1995) doubts the simplistic *r* vs. *g* contrast for economies with collective risk. Samuelson (1958), Diamond (1965) with capital, Tirole (1985) with a bubble, and most recently Blanchard (2019) have all generated *r < g* with Overlapping Generations (OLG).

This study is part of a rapidly developing stream of work on fiscal policy in circumstances when risk-free interest rates are persistently low, primarily in advanced economies. The view is divided: some support the notion that debt may not be so bad, in line with Blanchard (2019), while many explore various ways and argue there is little space for "free-lunch" (riskless) fiscal policies. Gopinath's remarks still hold: we do not know the answers yet (Blanchard et al., 2021).

Blanchard & Summers (2020) reminds part in a world where monetary policy cannot conduct stabilization policy (zero lower bound), fiscal policy has a critical role to play; hence there is no reason not to perform fiscal expansion and consequently increase public debt. Kocherlakota (2021) studies the public debt bubble and concludes that economic agents are still better off in the long run when more considerable debt and primary deficit policies are pursued.

Barrett (2018) found that government can safely borrow more due to the decline in nominal interest rates, but only by a few percentage points. Mehrotra & Sergeyev (2021) use an economical approach to debt sustainability, rather than flow (debt-to-GDP) and stock (r<g), explaining that so long as the primary surplus grows in lockstep with the level of debt, debt is sustainable. Mauro & Zhou (2021) explain that low differentials do not mean worriless borrowing. Maybe a little, but not really; costs can be volatile according to default histories. Cline (2021) said there is no fiscal free lunch forever. Lian et al. (2020) state that high public debt can make countries more vulnerable to an increase in growth differential, including low interest rates. Ball & Mankiw's (2021) findings conclude that fiscal policymakers should still be concerned about the crowding-out effects of government debt. Cochrane (2021) argues that large deficits still need to be repaid by subsequent surpluses, even when negative growth differentials occur.

Some literature dives deeper and explores the issue on the middle ground. Brumm et al. (2021) use simple closed and open economies, revisiting Blanchard's (2019) OLG model and conclude that when interest rates fall, risk-sharing by the government, rather than the budget deficit, may be the best way to achieve economic efficiency. Reis (2021) concludes that there is still a cap on how much the government may spend in a dynamically efficient economy with a bubble in the public debt. Policies can ease or tighten this cap through their independent effects on m – r and m – g. (m is the marginal product of capital). A. Mian et al. (2022) argue that debt might not blow up if $R < G - \varphi$ ($\varphi$ is R – G's sensitivity to debt) and the increase in deficits is modest ("free lunch"). That debt may not even arise if the economy is at the ZLB and the nominal growth rate is sufficiently responsive to increased deficits. Aguiar et al. (2021) find that a similar condition indicates the possibility of robust welfare improvements.

Little is known about the context of emerging economies. Several papers only include the data to form global dynamics (including Lian et al., 2020), while another explores the declining interest rate trend, with little link and analysis of public debt dynamics (Ruch, 2021). This study will contribute to this lack of literature by focusing on emerging economies and confirming whether Blanchard's conjecture still holds. His work with Subramanian (2021) has already done this, but only in the form of a policy brief and descriptive analysis. The result of this study can then decide whether the fiscal consensus is open for discussion or irrelevant to consider.

*2.3. Intergenerational Transfers and Welfare: Blanchard and Diamond*

The Overlapping Generation (OLG) model can describe the effects of a transfer from the young to the elderly and sound as the precursor to understanding the dynamics of debt and debt rollovers. Following Blanchard's (2019) replication of intergenerational analysis in Diamond (1965), we first explore intergenerational dynamics under the assumption of total certainty before loosening the restriction to a model with uncertainty.

Utility of people living for two periods; the first working and the second both consuming:

$$U = (1-\beta)U(C_1) + \beta U(C_2) \quad (1)$$

The budgetary constraints are as follows:

$$C_1 = W - K - D; C_2 = RK + D \quad (2)$$

$C_1$ is the consumption in the first period, while $C_2$ refers to the second period. $W$ stands for wage, $K$ stands for savings, $D$ stands for transfer from young to elderly, and $R$ is for the capital rate of return. In the setting of a constant returns production function, ignoring population increase and technological development, the growth rate becomes zero. For simplicity, labor is normalized to 1; hence $Y = F(K, 1)$. Both components are compensated based on the size of their marginal product. Utility maximization with a first-order condition:

$$(1-\beta)U'(C_1) = \beta R U'(C_2) \quad (3)$$

Change in utility as a result of a slight increase in the transfer $D$:

$$dU = [-(1-\beta)U'(C_1) + \beta U'(C_2)]dD + \\ [(1-\beta)U'(C_1)dW + \beta K U'(C_2)dR] \quad (4)$$

The first term (Equation 4) is subjected to a first-order condition (Eq. 3):

$$dU_a = [\beta(-RU'(C_2) + U'(C_2))] = \beta(1-R)U'(C_2)dD \quad (5)$$

The implication is relatively straightforward. When $R < 1$, a slight increase in the transfer enhances welfare. (ignoring other terms). This condition of dynamic inefficiency makes savers get a better rate of return by transfer rather than capital.

The second term (Eq. 4) demonstrates how debt affects utility by changes in $W$ and $R$. When we take on more debt, capital also declines, lowering wage and capital rate of return. Analyzing this matter further, specifically the welfare effect, one can use the factor price frontier relation $dR = F_{KK}(K, 1)dK$, therefore, $dR = -(1/K)dW$. Replace $dW$ in the second term (Eq. 4):

$$dU_b = -[(1 - \beta)U'(C_1) - \beta U'(C_2)]KdR \tag{6}$$

Utilizing first order condition (Eq. 3) for utility maximization:

$$dU_b = -[\beta(R - 1)U'(C_2)]KdR \tag{7}$$

A modest increase in the transfer boosts welfare when $R < 1$, similar to the first term. Interest rate increases as the result of lower capital stock. In the first period, a loss in capital leads to an equivalent fall in income, and in the second term (Eq. 4), an increase in income (given factor price frontier relation); more appealing compared to capital and accordingly enhances welfare.

Apply the relation between $dR$ and $dK$ and we get:

$$dU_b = [-\beta(R - 1)U'(C_2)]KF_{KK}(K, 1)dK \tag{8}$$

It is possible to rewrite the second term (Eq. 8) as:

$$dU_b = [\beta(1/\eta)\alpha][(R - 1)U'(C_2)]RdK \tag{9}$$

Two implications obtained from the analysis above: $R - 1$ determines the sign of the two effects. Utility increases from capital accumulation reduction when $R < 1$. The intergenerational transfer positively affects welfare in the steady state if the marginal product is smaller than the growth rate (in this case, 0). The second effect is determined by $\eta$ or substitution elasticity. The production function is linear when $\eta = \infty$. As a result, there is no impact on wage or rates of return on capital; null second effect.

Extension of the initial OLG is introduced by Blanchard (2019), which captures the uncertainty of capital's marginal product. When people are cautious, it stands to reason that the average safe rate will be lower than the marginal product of capital. The purpose of this extension is to understand which rate is of relevance for welfare purposes.

The expected utility of people born at time $t$:

$$U_t \equiv (1 - \beta)U(C_{1,t}) + \beta EU(C_{2,t+1}) \tag{10}$$

Budgetary constraints:

$$C_{1t} = W_t - K_t - D; C_{2t+1} = R_{t+1}K_t + D \tag{11}$$

Production function (Constant Returns; $N = 1$ and stochastic $A_t$):

$$Y_t = A_t F(K_{t-1}, N) \tag{12}$$

Capital at time $t$ represents the young generation's savings at time $t - 1$.

Utility maximization with the first-order condition at time $t$:

$$(1 - \beta)U'(C_{1,t}) = \beta E[R_{t+1}U'(C_{2,t+1})] \tag{13}$$

Based on Equation 13, the shadow safe rate can now be defined as $R^f_{t+1}$ fulfilling:

$$R^f_{t+1} E[U'(C_{2,t+1})] = E[R_{t+1}U'(C_{2,t+1})] \tag{14}$$

Consider the impact of a slight increase in $D$ on utility at time $t$:

$$dU_t = -(1 - \beta)U'(C_{1,t}) + \beta EU'(C_{2,t+1})]dD \tag{15}$$
$$+[(1 - \beta)U'(C_{1,t})dW_t + \beta K_t E[U'(C_{2,t+1})dR_{t+1}]$$

The partial equilibrium and direct effect of transfer are represented by the first term ($dU_{at}$), which is similar to the basic OLG model. The second term ($dU_{bt}$) describes the general equilibrium and indirect influence of wage and rate of return on capital changes. The first term is subjected to a first-order condition (Eq. 13) and using the definition of the safe rate (Eq. 14):

$$dU_{at} = \beta(1 - R^f_{t+1})EU'(C_{2,t+1})dD \tag{16}$$

The relevant rate, which determines the sign effect of the transfer on welfare, is the <u>safe rate</u> for the first channel. $R^f_{t+1} < 1$ means that the transfer is welfare improving. An explanation behind this would be that when compared to the risk-adjusted rate of return on capital (safe rate), the intergenerational transfer offers a greater rate of return.

First order condition (Eq. 13) implemented in the second term (Eq. 15):

$$dU_{bt} = \beta E[R_{t+1}U'(C_{2t+1})]dW_t + \beta K_t E[U'(C_{2,t+1})dR_{t+1}] \tag{17}$$

Consider the equation:

$$dW_t = -K_{t-1}dR_t \tag{18}$$

Replacing $dW_t$ (using Eq. 18) and using relations between $dR_t$ and $R_t$ and between $dR_{t+1}$ and $R_{t+1}$ in the $dU_{bt}$ equation (Eq. 17):

$$dU_{bt} = [-\beta \frac{KF_{KK}(K)}{F_K(K)} E[U'(C_{2t+1})R_{t+1}]](R_t - 1)dK \tag{19}$$

Because $KF_{KK}(K)/F_K(K) = -(1/\eta)\alpha$ and we have the shadow safe rate relation, rewrite the $dU_{bt}$ expression (Eq. 19) as:

$$dU_{bt} = [\beta(1/\eta)\alpha]E\left[\left(R_{t+1} - \frac{R_{t+1}}{R_t}\right)U'(C_{2,t+1})\right]R_t dK \tag{20}$$

The relevant rate is the <u>risky rate</u> (marginal product of capital) for the second channel. $R_t < 1$ indicates that the implicit transfer resulting from a change in input prices raises utility; the contrary happens when $R_t > 1$. Since capital delivers $R_{t+1}$, decrease in the capital, which results in price changes, reflects implicit transfer with rate of return $R_{t+1}/R_t$; less if $R_t > 1$.

Combining the two sets of results: when the safe rate is less than one, and the risky rate is more than one, the two terms work in opposing directions. The first term implies a positive effect of debt towards welfare, while the second implies a negative effect. Both rates are indeed relevant.

To gain a notion of the relative magnitudes of the two effects, we may assess the two terms on average values of safe and risky rates:

$$dU/dD = [(1 - ER^f) - (1/\eta)\alpha ER^f(-dK/dD)]\beta E[U'(C_2)] \qquad (21)$$

$$signdU \equiv sign[(1 - ER^f) - (1/\eta)\alpha ER^f(-dK/dD)(ER - 1)] \qquad (22)$$

Approximation to $dK/dD$ can be derived from the accumulation equation. Ignore uncertainty and the small wealth effect of the transfer on saving so that capital in a steady state fulfills:

$$K = \beta W - D = \beta F_N(K, 1) - D \qquad (23)$$

Equation 23 implies that for a small transfer, $dD$:

$$dK = \beta F_{KN}(K, 1) - dD \qquad (24)$$

Rewrite Equation 24 using the elasticity of substitution and labor share definitions:

$$dK/dD \approx -\frac{1}{1 - \beta\alpha(1/\eta)ER} \qquad (25)$$

If the production is linear ($\eta = \infty$), the second term becomes 0, and $ER^f$ is the only rate that matters. If $ER^f < 1$, higher transfer leads to higher welfare. The price effect is strengthened by lower substitution elasticity, and the welfare impact ultimately reverses and becomes negative. The Cobb Douglas case ($\eta = 1$) using the approximation $ER \approx (1 - \alpha)/(\alpha\beta)$ can be simplified:

$$signdU \equiv sign[(1 - ER^f ER)] \qquad (26)$$

$sign$ implies the result showing between positive or adverse effects. According to the analysis, a transfer's welfare effects may not necessarily be harmful, or if they are, they may not be significant.

## 3. METHODOLOGY

The methodology used in this study will be in the form of model simulations based on the work of Acalin (2019), which fully replicates the analysis of the stochastic overlapping generations (OLG) model developed by Blanchard (2019) within the structure of the model presented on the theoretical foundation. The models are slightly modified with different parameters compared to the original to fully reflect the proposed dynamics in a different context: Indonesia as one of the emerging market countries.

Each life period (from two periods) is set to 25 years. Because risk aversion determines the difference between average safe and risky rates, having a different elasticity of substitution between two life periods and a risk aversion degree is desirable. For this purpose, utility is assumed to have an Epstein-Zin-Weil (Epstein & Zin, 2012; Weil, 1990) form representation:

$$(1-\beta)lnC_{1,t} + \beta\frac{1}{1-\gamma}lnE\big(C_{2,t+1}^{1-\gamma}\big) \qquad (27)$$

The intertemporal elasticity of substitution equals one; hence the log-log specification, and $\gamma$ is the coefficient of relative risk aversion. The production function is as follows:

$$Y_t = A_t\big(bK_{t-1}^{\rho} + (1-b)N^{\rho}\big)^{1/\rho} = A_t\big(bK_{t-1}^{\rho} + (1-b)\big)^{1/\rho} \qquad (28)$$

$A_t$ is white noise with log-normal distribution $ln\,A_t \sim \mathcal{N}(\mu;\sigma^2)$ and $\rho = (\eta-1)/\eta$. $\eta$ represents the substitution elasticity. The substitution elasticity determines the second effect's strength; therefore, production is assumed to be a constant elasticity with multiplicative uncertainty. A linear production function is present when $\eta = \infty$ and $\rho = 1$. Cobb Douglas is the limit when $\eta$ approaches 1; thus $\rho$ approaches 0. Definitions of wage and rate of return:

$$W_t = A_t(1-b)\left(\frac{Y_t}{A_tN_t}\right)^{1-\rho} = A_t(1-b)\left(\frac{Y_t}{A_t}\right)^{1-\rho} \equiv W(K_{t-1},A_t) \qquad (29)$$

$$R_t = A_tb\left(\frac{Y_t}{A_tK_{t-1}}\right)^{1-\rho} \equiv R(K_{t-1},A_t)\,;R_{t+1} = A_{t+1}b\left(\frac{Y_{t+1}}{A_{t+1}K_t}\right)^{1-\rho} \equiv R(K_t,A_{t+1}) \qquad (30)$$

The wage has a log-normal distribution, which may lead to arbitrarily small numbers. The model integrates the reception of non-stochastic endowment to ensure that deterministic transfer from the young and elderly is always possible, denoted by $X$, for the young with the amount of 100 percent average wage (without including transfer) for each pair $ER, ER^f$.

Based on stylized facts established per Indonesia's state, the parameters are chosen to fit a set of values for the net annual average risky rates minus the growth rate and net annual average safe rates. Safe rate is the risk-free rate, the cost of funds for the government to conduct public debt. Country-based 10-Year Bond Yield is used as a reference. Risky rate is defined as the average marginal product of capital, which in this study is based on lending rate: Commercial banks' weighted average rate on working capital loans to the private sector in national currency. The amount of the loan weights the rate.

Other coefficients are chosen following Blanchard's (2019) a priori choices, with possible modifications for analyzing various situations. Cobb-Douglas Capital share, b, is equal to 1/3. $\sigma$ is equal to 0.2. Two values of $\eta$: $\eta = \infty$ for the linear production and $\eta = 1$ for the Cobb-Douglas. Fitting each pair of $ER, ER^f$ uses $\beta$ and $\mu$ on the one hand and $\gamma$ on the other. $\beta$ and $\mu$ determine the average capital accumulation rate and average capital marginal product (risky rate). Generally, both factors are essential, but one should note the dynamics in extreme cases: linear and Cobb-Douglas, used in the analysis.

Linear production only depends on $\mu$ since the marginal product of capital is unaffected by capital levels. The opposite is true for Cobb-Douglas, which only depends on $\beta$. For both cases, only the applicable parameter is selected to suit the relevant average value of the capital's marginal product. $\gamma$ and $\sigma$ establish the spread between risky and safe rates. The spread determines $\gamma$. Relation between safe and risky rate (without transfers present):

$$lnR^f_{t+1} - lnER_{t+1} = -\gamma\sigma^2 \qquad (31)$$

Three groups of equations are solved and simulated. First, equations of motion without intergenerational transfers or debt. They are used to solve for $\beta, \mu, \gamma$, which generate $ER, ER^f$ and subsequently, simulation based on 30,000 $A$ realizations is utilized to specify the average level of capital used as starting value of capital when transfers or debt are introduced. Second, equations of motion with intergenerational transfers. In this part, the equations solve the change of steady state welfare without and with intergenerational transfers for each pair $ER, ER^f$. This is done by generating a sequence of 2,000 realizations of $A$, then computing relevant averages. Third, equations of motion with a debt rollover. They derive dynamics of debt to saving ratio and utility by 1000 simulations, each based on five draws of $A$ (150 years period). The simulation starts from the average steady-state value of capital without debt, then introduces debt at time 0.

Debt dynamics can be analyzed by looking at debt behaviors based on the simulations. Debt rollovers can either be sustainable or not, and deciding whether a particular debt rollover fails will require establishing a threshold. In this study, the arbitrary setup of Acalin (2019) is followed, which sets a slightly different condition compared to Blanchard (2019), that debt exceeded 115% of the initial increase in debt as the threshold. From that point forward, debt dynamics are projected to result in a sustained increase in the debt-to-saving ratio. Another assumption is that if the threshold is exceeded, debt is fully paid off by taxing the young generation. Despite this setting, the results will still be similar to the original paper.

## 4. RESULTS

*4.1. Indonesia's Interest Rate-Growth Differentials*

Analyzing the trajectories of Indonesia's interest and growth differentials (*r-g*), fundamental in deciding how debt can theoretically be sustainable given specific public debt levels and the scope of adjustment, requires three data points: nominal interest rate, nominal growth rate, and lending rate. The nominal interest and growth rates will determine the growth differentials based on the safe rate (risk-free government lending rate). In contrast, the lending and the same nominal growth rates will establish risky growth differentials as a comparison. This is in line with the concept of debt under uncertainty discussed in the theoretical foundation.

The ideal way is to use as many data points as possible to analyze broader dynamics. While the data available for nominal growth rate and lending rate is quite comprehensive[2], nominal interest rate based on Indonesia's 10-Year Bond Yield is only available from August 2004, making it the starting point of the observation of the dynamics.

Safe rate interest rate-growth differentials are almost always negative, except in 2019 when it is only slightly above zero. On the components, both the nominal interest rate and nominal growth rate tend to decline, in accordance with many findings of the declining global trend of interest rates. Moving towards risky rates, the results are mixed. Starting with another slightly above zero differentials in 2004, the differentials are generally negative in the 2005-2011 period, except 2009, possibly due to the Global Financial Crisis (GFC). From 2012 onwards, differentials become positive with no clear-cut trend on whether this would stay positive and increase as time goes on or back to negative.

Based on key statistics in both cases, a new range of parameters is devised to simulate Indonesia's public debt dynamics based on the framework developed by Blanchard (2019).

**Table 4.1 Indonesia's Interest Rate-Growth Differentials Key Statistics**

| SAFE | | RISKY | |
|---|---|---|---|
| Median | -2.82 | Median | 1.03 |
| Mean | -4.75 | Mean | -1.12 |
| Max | 0.66 | Max | 4.29 |
| Min | -14.10 | Min | -11.66 |
| Q1 | -7.62 | Q1 | -4.40 |
| Q3 | -1.91 | Q3 | 1.81 |

---

[2] Starting from Year 1960 and 1986 respectively, collected from World Bank's WDI for the growth rate and International Monetary Fund's International Financial Statistics for the lending rate. International Financial Statistics received sources from Bank Indonesia and Ministry of Finance Statistics Department.

The Blanchard model sets the parameter range of two main variables as reference points in simulation: safe rate (Rf) and risky rate (MPK, capital marginal product). Given the evolution of the US growth rate and interest rate, Blanchard calibrates the model to suit a set of average safe and risky rate values: Average net annual risky rates (marginal products of capital) minus growth rate of 0 percent to 4 percent, and average net annual safe rates minus growth rate of 2 percent to 1 percent.

This study calibrates the model under a similar approach with a different set of values based on the findings above: risky differentials between 1 percent and 5 percent and safe differentials between -3 percent and 2 percent. Despite maximum data values of 0.66 and 4.29, 1 additional percentage point is added arbitrarily to adjust to Indonesia's higher volatility risk, considering its status as a developing country.

*4.2. Stochastic Overlapping Generations (OLG) with Global and Indonesia Growth Differentials*

Blanchard's (2019) analysis of the global low interest rate-growth differentials is compared with an analysis using the same framework but considering different parameters based on Indonesia's data. Running python codes within Acalin's (2019) Jupyter Notebook shows comparable results by simulating the model under various scenarios.

Transfers Welfare Implications

*Linear Production Function*

In the case of linear production, Figure 1 shows the effects of a small transfer (5 percent of pre-transfer average saving) on welfare for various combinations of the safe and risky rates (reported as net rates at annual values rather than as gross rates at 25-year values for convenience). No matter the average risky rate, welfare rises if the safe rate is negative (more precisely, if it is less than the growth rate, which in this case is equal to 0). Figure 2 shows a higher transfer (20 percent of savings) in the linear production scenario. A higher ER leads to a smaller utility boost if welfare rises and a larger welfare reduction if welfare falls for a given ERf.

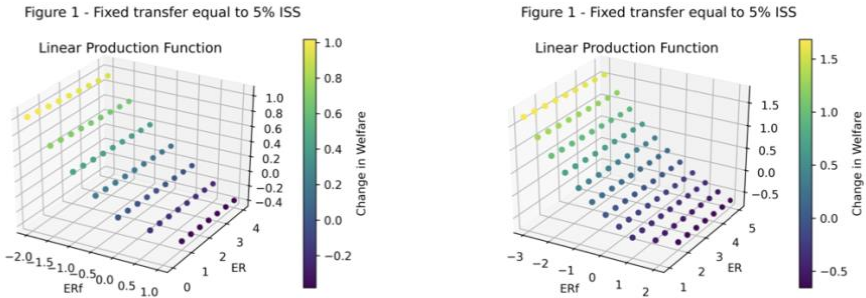

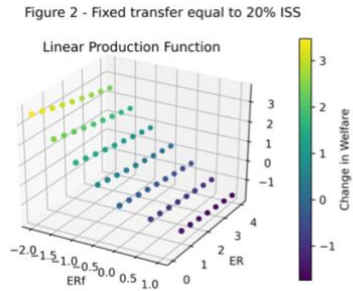 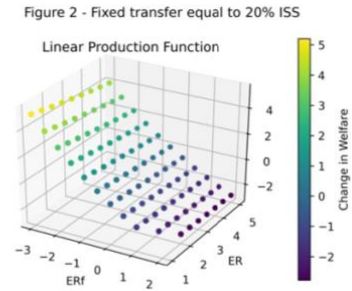

Under the modified model, the dynamics stay the same for both the effects of a small transfer and a larger transfer (5 percent and 20 percent of average savings before the transfer, respectively.). With a higher possibility of lower differentials, the change in welfare can be much more significant. The graph shows the possible change in welfare from a maximum of 1.0 to 1.5 (5 percent) and 3 to almost 5 (20 percent). Indeed the greater the transfer, the higher the change in welfare. The caveat found when comparing the original and modified model is that when the transfer is higher, there is also a higher overall possibility of lower change in welfare, as shown in the modified model (20%) with more than -2 change. This risk is not found in the 5% linear fixed transfer.

*Cobb-Douglas Production Function*

Figures 3 and 4 accomplish the same thing, but this time for the Cobb-Douglas scenario. Both impacts are in play right now, and both rates are essential. A lower safe rate means the transfer is more likely to improve welfare; a greater risky rate means it is less probable. So long as the risky rate is less than 2 percent above the growth rate, a safe rate 2 percent lower than the growth rate leads to a gain in welfare for a modest transfer (5 percent of saving). As long as the risky rate is less than 1 percent higher than the growth rate, a safe rate 1 percent lower than the growth rate increases welfare.

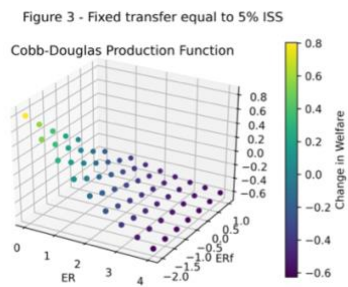 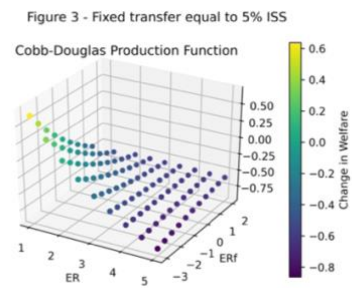

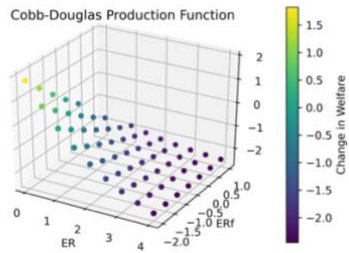
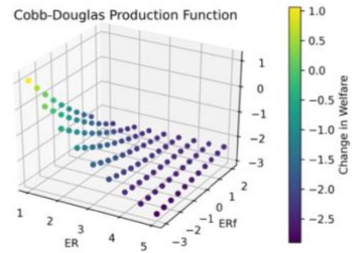

Cobb-Douglas Production Function produces contrasting dynamics compared to the linear approach. While the insight that can be drawn from the linear function, at a glance, is that higher risk offers better returns, different parameters of the modified model instead only yields a higher risk of high negative change in welfare (0.8 from 0.6 in 5%, more than 2.5 from 2 in 20%). The maximum change in welfare goes down to only 0.6 from 0.8 (5%) and 1 from 1.5 (20%), a stark contrast to linear simulation.

<u>Debt Rollovers Welfare Implications</u>

*Debt Dynamics*

The government issues debt in the amount of D0. After the original issuance and related transfer, there are no taxes or subsidies until the debt rollover fails. Figure 6 depicts 1,000 random debt evolutions paths, assuming the production function is linear. Figure 8 plots 1,000 stochastic paths of debt evolutions under the assumption that the production function is Cobb-Douglas. In both scenarios, the initial debt rise equals 15% of the average steady-state saving (pre-debt). In both scenarios, the underlying parameters are calibrated to fit values of ER and ERf absent debt of 1% for the yearly safe rate and 2% for the annual risky rate, respectively. In the context of Indonesia, the underlying parameters are adjusted to -0.5 percent for the annual safe rate and 3% for the annual risky rate, following the midpoint of the safe and risky parameter ranges.

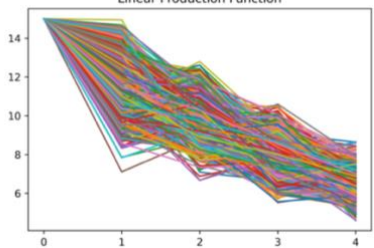
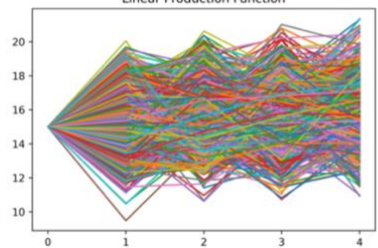

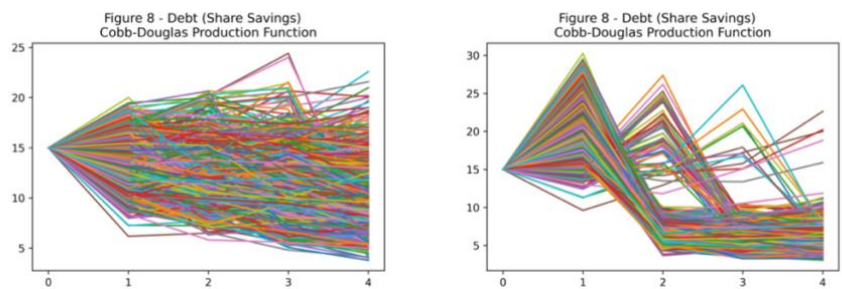

The point at which the debt gets sufficiently large and positive is defined as failure (so that the probability that it does not explode becomes very small, depending on the unlikely realization of successive large positive shocks which would take the safe rate back below the growth rate). Acalin (2019) arbitrarily set the threshold at 115 percent of the initial debt rise. If the debt rollover fails, it is assumed that a portion of the debt will be repaid by a tax on the young, bringing the debt level back to 40% of the original debt increase. This exaggerates the impact of failure on the young at that period, yet it is the most straightforward to depict.

*Welfare*

Debt rollovers are far less appealing than a pay-as-you-go plan. Keep in mind the intergenerational transfer's two implications. The first stems from the fact that people earn a 1 percent rate of return on their transfer, which is often higher than Rf. They receive a rate of return of only Rf in a debt rollover, often less than 1. They are unconcerned with holding debt or capital on the margin. There is still an inframarginal impact, a consumer surplus (in the form of a less risky portfolio, and hence less dangerous second-period consumption). However, it has a lower positive effect on welfare than the simple transfer system. The second effect, attributable to changes in wages and rates of return on capital, is still present. The net effect on welfare is more likely negative, even if it is less persistent as debt declines over time.

Figures 5 and 7 exhibit the average welfare impacts of successful and unsuccessful debt rollovers in the linear and Cobb-Douglas settings. Debt rollovers rarely fail in the linear scenario, and welfare increases over time. The effect is positive and considerable for the generation that receives the initial transfer linked with debt issuance. While later generations are mainly agnostic about whether they should invest in safe debt or risky capital, the inframarginal benefits (from a less risky portfolio) imply marginally higher utility. However, compared to the initial welfare effect on the elderly from the initial transfer, the welfare increase is minimal (about 0.18 percent at first and diminishing over time) (8.75 percent).

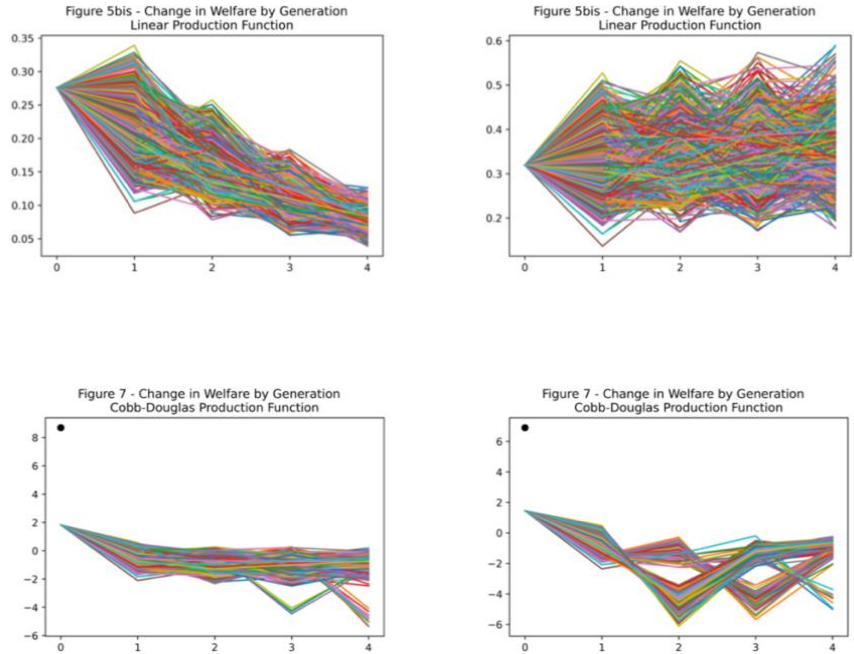

However, in the Cobb-Douglas scenario, this positive effect is more than compensated by the price effect. While welfare rises for the first generation (by 2%), it often declines after that. The average adverse welfare cost lowers as debt diminishes over time in the case of successful debt rollovers. In the event of failed rollovers, the adjustment entails a greater welfare loss when it occurs.

Insights from the modified model are different. Under linear production assumption, debt rollovers are unpredictable, not having a clear pattern of success and failure of debt dynamics. Welfare received is also the same, with a lower initial welfare effect on the old (7 compared to 8.75) but unpredictable welfare gain trends over time. It is not decreasing, and it is unknown how it will turn out.

Similar to the linear case, the initial change in welfare for the Cobb-Douglas modified model is lower than the original one. First-generation welfare gain is the same, but the rest is quite tricky to analyze: also typically negative, but the dynamic varies. There are many scenarios where on the third generation the change goes towards positive, but others fall up to -6 level, and then those which fell goes up again, while the upward goes down on the fourth generation. Some, but not as much as the preceding two dynamics, go slightly upwards until the fourth generation and then tumble down. The change is unpredictable and more extreme.

Another interesting result to dive into is the debt rollover dynamics on Cobb-Douglas. Moving towards the second generation, debt can rise to 30% from the initial point of 15%, much higher than 20% in the original simulation. When debt is lower (under some scenarios), the minimum point is also a bit high, 10% compared to 5%. After that, the trends diverge. Some scenarios peaked at the third generation, others fourth or continue to rise. Most scenarios, however, stabilize at a lower debt share.

All in all, the modified model generally shows that the debt dynamics get worse when adjusting to Indonesia's interest-rate growth differentials. Increasing public debt will lead to much more volatile, risky outcomes with high uncertainty.

*4.3. Changing Parameters*

This section explores further adjustments to put into context Indonesia's case of debt dynamics using Blanchard's (2019) model for simulation. For both the original and modified model, four different factors are adjusted. Share of government spending allocated to the old generation, capital share in Cobb-Douglas, initial non-stochastic endowment, and upper debt limit (debt threshold) for default.

First is the initial non-stochastic endowment. Given that wages follow a log-normal distribution and can thus be arbitrarily small, the endowment $X$ is required to ensure that deterministic transfers from the young to the elderly are always possible, regardless of W's realization. The original model assumed that the endowment would be equal to 100% of the average wage if the transfer were not made. However, this assumption is unrealistic; therefore, the initial endowment is modified to 75 percent of the wage.

Overall, almost no change in intergenerational transfers and debt rollovers in the linear production function. The difference can be seen in the Cobb-Douglas debt rollovers, where relaxing the assumption of initial endowment reduce the change in welfare. More paths lead to high adverse effects on welfare and higher debt share, , tantamount to Evans's (2020) findings in replicating Blanchard's (2019) model and modifying several assumptions. The same conclusion holds when observing modified model outcomes. Despite several outliers with much lower debt shares, the overall change in welfare worsened, and debt share went up in linear and Cobb-Douglas cases.

Second, capital share in the Cobb-Douglas. Blanchard (2019) choose the coefficient a priori 1/3. This study modifies the coefficient to 0.5 instead. In both fixed transfer scenarios of intergenerational transfer, increasing the capital share also increases the variance of change in welfare. On the original model, the range goes from [-0.6, 0.8] to [-1, 1.5] and [-2, 1.5] to [-4, 4], while on the modified model, it goes from [-0.8, 0.6] to [-1, 1.5] and [-2.5, 1] to [-4, 3]. Debt rollovers change only slightly, with a higher possibility of scenarios having negative changes in welfare in later generations and less initial change in the first generation (on the original).

Debt share differs between models: the original one stays the same, while the modified model shows that the change in capital share increases the chance of debt share escalation.

<u>Third</u>, the share of government spending allocated to the old generation. To compare, the share is toned down to 0.9. This change of parameter does not change any of the debt dynamics. <u>Lastly</u>, the threshold of the initial increase in debt. This parameter is adjusted to 110 percent from 115 percent, making it even more strict for debt explosion avoidance. Looking at the original model, a change in threshold does not change the dynamics dramatically, only touching the Cobb-Douglas debt rollovers with more scenarios on higher negative change in welfare territory and little tightening in debt share range.

Nevertheless, debt rollovers in the modified model Cobb-Douglas show peculiar outcomes. Trajectories of linear production now have negative scenarios of change in welfare, trading off with lower debt share. Cobb-Douglas now has a single trend both in welfare change and debt share. Initially, the change in welfare is negative, then slowly moves to above zero zone. Debt share goes down in all scenarios (with different scales of decline), though some show possible dynamics back to the upward trend.

*4.4. Discussion*

In general, each change in the four factors explained above gives insights into the analysis of debt dynamics. Relaxing the assumption of initial endowment lessens the desirability of increasing public debt, both in the form of transfers or debt rollovers. When the economy has a higher capital share, as reflected in the Cobb-Douglas model, change in welfare in implementing intergenerational transfer varies much more; hence while the welfare can increase more than usual, it can also have serious repercussions. Debt rollovers are also worsening in welfare change and, in the case of the modified model, higher debt share. Change in government spending allocation to the old generation does not affect debt dynamics.

On the contrary, a change in debt threshold to default also reduces welfare while reducing debt share variance (a trade-off). Specifically, on the linear production modified model, which shows Indonesia's possible dynamics, the trade-off between debt share and welfare change is present. Cobb-Douglas is similar with a dramatic change in trend compared to before changing the parameter.

Closer to reality and implementing the OLG framework in Indonesia's case, the finding is quite apparent. Increasing public debt may have dire consequences since Blanchard's (2019) explanation of its advantages does not seem to hold after simulating the same tools with minor modifications. Debt share and welfare trade-offs also exist, making the possibility of debt explosion or even a slight rise much less desirable since reversing the trend may eventually reduce the welfare of the society—offsetting the purpose in the first place.

## 5. CONCLUDING REMARKS

The trend of negative interest rate-growth differentials opens up the discussion about how higher public debt may not be costly. Simple debt arithmetics show that when the ability to pay the debt, that is the economy's growth, is higher than the cost of capital, debt rollovers shall still be successful even when raising public debt; the opportunity is there. This study contributes to the discussion by simulating OLG models introduced by Blanchard (2019) under uncertainty, showing debt and welfare dynamics in two scenarios: intergenerational transfers and debt rollovers in the case of Indonesia. This is done by modifying the model parameters based on interest rate-growth differentials historic data from 2004-2019, which results can be compared to the original model simulations to show a different perspective of emerging economies, specifically Indonesia, and a conclusion can be made on the ability to make use of the negative interest rate-growth differentials.

Comparing the results of both the original and modified model, it can be seen that the fiscal consensus does not hold when implementing Blanchard's (2019) analysis with Indonesian-based rate parameters. Increasing public debt makes the economy more volatile, uncertain, and high-risk results. Adding into perspective the modification of other factors supports the initial finding even more; a lower initial endowment for each individual also diminishes the benefits of public debt and higher capital share under the Cobb-Douglas production function. When the risk of debt explosion appears, that is when the simulation gets to the point of the debt threshold set by the model, a trade-off between welfare and debt share exists—efforts to reduce debt share will reduce the welfare of the society; hence higher public debt should be considered with caution.

The policy implication of the study is quite clear: be careful of the opportunity. At first, the fiscal consensus seems to make sense with the availability of negative interest rate-growth differentials. However, the higher volatility of a developing economy and uncertain future of the trend shows otherwise. Increasing public debt may not be the way to go, avoiding possible dire consequences. Another factor to consider is the current developments of interest rates, which seem to be contrary to many research findings: reverse of the decreasing trend. Should this be the case, the opportunity of negative interest rate-growth differentials will not hold, implying it is better to avoid raising the public debt substantially. This study is subject to several limitations with the simplicity of the model. Assumptions are unrealistic, and the model does not portray the ideal depiction of the economy. The data is limited; no government bond yield can be analyzed before the 21st century. These limitations are deeply considered, but this study has its boundaries: time, costs, and efforts.

Further research should be conducted to confirm the model exploration findings and compare it with other emerging economies. Future research agenda may include a direct comparison between similar countries with Indonesia's profile as an emerging market, an extension of the available data using other comparable indicators, exploring the limitations of the OLG model to portray debt dynamics in various cases, and multiple equilibria considering the possibility of long-run rising trend of interest rate. The model can also be modified to make it more realistic and fit the profile; hence the results are ideal to conclude and serve as a reference for policy-making. The possibility of interacting with many global and local trends, such as income inequality and demographic bonus, is also interesting.